\documentclass[submission,copyright,creativecommons]{eptcs}
\providecommand{\event}{QAPL 2017} 
\usepackage{breakurl}             
\usepackage{xspace} 
\usepackage{graphicx} 
\usepackage{microtype} 
\usepackage{placeins} 
\usepackage{amsmath}
\usepackage{stmaryrd}
\usepackage{amssymb}
\usepackage{tikz}
\usetikzlibrary{shapes,arrows} 
\usetikzlibrary{positioning}
\usetikzlibrary{chains}
\usetikzlibrary{automata}
\usetikzlibrary{patterns}
\usetikzlibrary{decorations.pathreplacing}
\usetikzlibrary{calc,decorations.markings,decorations.pathreplacing,fit,backgrounds,shapes.symbols,shapes.geometric}
\tikzset{
  gnode/.style={draw,shape=circle,inner sep=0,minimum height=.2cm,
    minimum width=.2cm},
  hyperedge/.style={shape=rectangle,draw,inner sep=0,minimum width=.6cm,
    minimum height=.4cm}
}
 
\tikzset{every fit/.style={shape=rectangle,inner sep=5pt}}
\newtheorem{defi}{Definition}
\tikzset{
  mono/.style={>->},  
  ontop/.style={preaction={draw,-,line width=3pt,white}},
  arlab/.style={circle,inner sep=1pt,font=\scriptsize}
}

\usepackage[colorinlistoftodos,
textsize=footnotesize,color=orange!70]{todonotes}

\newcommand{\fonta}[1]{\textsl{#1}}

\title{\textsc{Paws}: A Tool for the Analysis of Weighted Systems\thanks{Research partially supported by DFG project BEMEGA.}}
\author{Barbara K\"onig
\institute{Universit\"at Duisburg-Essen}
\email{barbara\_koenig@uni-due.de}
\and Sebastian K\"upper
\institute{Universit\"at Duisburg-Essen}
\email{sebastian.kuepper@uni-due.de}
\and Christina Mika
\institute{Universit\"at Duisburg-Essen}
\email{christine.mika@uni-due.de}
}
\def\titlerunning{Paws: Program for the Analysis of Weighted
  Systems}
\def\authorrunning{B. K\"onig, S. K\"upper \& C. Mika}
\begin{document}
\maketitle
\newcommand{\paws}{\textsc{Paws}\xspace}
\renewcommand{\phi}{\varphi}

\begin{abstract}
  \paws is a tool to analyse the behaviour of weighted automata and
  conditional transition systems. At its core \paws is based on a
  generic implementation of algorithms for checking language
  equivalence in weighted automata and bisimulation in conditional
  transition systems. This architecture allows for the use of
  arbitrary user-defined semirings. New semirings can be generated
  during run-time and the user can rely on numerous automatisation
  techniques to create new semiring structures for \paws'
  algorithms. Basic semirings such as distributive complete lattices
  and fields of fractions can be defined by specifying few parameters,
  more exotic semirings can be generated from other semirings or
  defined from scratch using a built-in semiring generator. In the
  most general case, users can define new semirings by programming (in
  C\#) the base operations of the semiring and a procedure to solve
  linear equations and use their newly generated semiring in the
  analysis tools that \paws offers.
\end{abstract}

\section{Introduction}

In recent times, modelling techniques have shifted from a purely
acceptance-based reasoning to one that takes various notions of weight
and quantities into consideration. The theory of weighted automata
gives rise to a flexible framework, where acceptance behaviour can be
quantified in numerous ways. Probabilities are commonly used to
express the likelihood of making a given transition, whereas the
tropical semiring is often used to denote the cost of making a
transition. Due to the great variety of semirings the theory of
weighted automata can be applied to a multitude of different fields of
interest, such as natural language processing, biology or performance
modelling (see e.g. \cite{hwaDKV}). Typical questions regarding
weighted automata concern their language (or traces), for instance
language equivalence. Due to the generality of the notion of weighted
automata, they are a versatile means for modelling purposes, though
decidability results are not as strong as for non-deterministic
automata. In particular, it is well-known that language equivalence is
an undecidable problem for weighted automata over the tropical
semiring \cite{Krob94theequality}. This, however, is not a damning
result for language equivalence, since it is also well-known that
language equivalence is decidable for many other semirings such as the
two-valued Boolean algebra (for which weighted automata are just
nondeterministic finite automata) or fields (where decision procedures
run in polynomial time
\cite{kmoww:language-equ-prob,b:weighted-bisimulation}). \paws
\footnote{The tool can be downloaded from
  \texttt{www.ti.inf.uni-due.de/research/tools/paws/}} is a tool that
offers algorithms to decide language equivalence and the threshold or
universality problem for weighted automata.

Based on our previous work in \cite{KK16} and
\cite{bkk:up-to-weighted}, the tool \paws implements two different
approaches to language equivalence checks for weighted automata. One
approach employs a backwards search assuming at first that all states
are language equivalent and exploring words from the end to the
beginning to determine non-equivalent pairs of states, similar to
partition refinement. Different from usual partition refinement
algorithms, this algorithm does not necessarily terminate at the
moment when the partitions cannot be further refined in one step,
because refinements can happen at a later point of time. The
termination condition has to be chosen differently here, and as
expected, the algorithm does not necessarily terminate for all
semirings (see the undecidability result by Krob
\cite{Krob94theequality}). However, for many semirings it does and in
either case it is always a suitable semi-decision procedure for the
absence of language equivalence. This approach is closely related to
the line of work started by Sch\"utzenberger,
\cite{DBLP:journals/iandc/Schutzenberger61b} and later generalised in
\cite{bls:conjugacy}, using the notion of conjugacy.

Additionally, \paws offers a second approach to decide language
equivalence \cite{bkk:up-to-weighted}, which stands in the tradition
of Bonchi's and Pous' seminal work on equivalence checks for NFAs
using up-to techniques \cite{bp:checking-nfa-equiv}.  Here, a language
equivalence relation on vectors is built, starting from the initial
pair of vectors suspected to be language equivalent. As the algorithm
progresses, it builds a relation on vectors similar to a bisimulation
relation and stops at the moment when a suitable relation proving
language equivalence is found, or a witness to the contrary
appears. The algorithm works up-to congruence and therefore prunes the
relation on-the-fly, dropping redundant vector pairs. The flexibility
of this optimised variant of the algorithm is reduced when compared to
the partition refinement algorithm, but it can still be used for a
variety of user-generated semirings, such as rings and $l$-monoids,
and can lead to an exponential speed up in some cases. Based on a
similar approach, \paws also offers a decision procedure for the
universality problem for weighted automata over the tropical semiring
of the natural numbers, which is potentially exponentially faster than
a naive approach due to Kupferman et
al. \cite{abk:decidable-weighted-automata}. The universality problem
checks whether from a given initial vector, all words have a weight
smaller than or equal to a given threshold.

Finally, \paws also considers conditional transition systems, which,
rather than adding weights to transitions, extend traditional
transition systems by means of conditions, or product versions to
enable flexible modelling of software product lines, while taking
possible upgrades between different products of a software product
line into account \cite{DBLP:conf/icse/CordyCPSHL12}. For these kinds
of systems, bisimulation rather than language equivalence is
considered, because the user experience for different products is in
the focus. The bisimulation check can be performed on any finite
distributive lattice defined via its set of join irreducible
elements. Alternatively, a BDD-based implementation of (certain)
lattices is offered and allows for a significantly faster bisimulation
check as presented in \cite{bkks:cts-upgrades}. Again, the
bisimulation check is parametrised over the lattice used and can
accept any lattice, either defined directly via its irreducible
elements, or using the BDD-based approach, as input.

A key feature of \paws is its extensibility. The algorithms are
parametrised over the semiring and it is therefore possible to use the
algorithms \paws offers, not only for the semirings that come
pre-implemented, but also for newly generated semirings. For that
purpose, \paws offers ways of adding new semirings and executing
algorithms for these semirings. All algorithms are implemented
generically and can be used for various semirings or $l$-monoids,
provided all necessary operations such as addition, multiplication,
and solving linear equations are specified. Therefore, \paws is
equipped with a semiring generator that allows to generate new
semirings that are not pre-implemented and to define weighted automata
or conditional transition systems over these. Specifically, we have
built several layers of automatisation, so that whenever one, e.g.,
defines a complete distributive lattice, it suffices to give a partial
order, from which the lattice of downward-closed sets is generated
\cite{dp:lattices-order} and all operations are provided
automatically. Building semirings from other semirings using
crossproducts is almost completely automatised and modulo rings are
automatised using Hensel liftings \cite{Hensel:lifting} for solving
linear equation systems. In addition, it is possible to add arbitrary
semirings by providing code for the operations mentioned before.

\section{Preliminaries: System Types and Decision Problems}
\label{sec:preliminaries}

Here we give a short overview over the systems that \paws can analyse,
i.e. weighted automata and conditional transition systems. For that
purpose we require the notions of semirings and distributive lattices.

\begin{itemize}
\item A \emph{semiring} is a tuple $\mathbb S=(S,+, \cdot, 0, 1)$
  where $(S,+,0)$ is a commutative monoid, $(S,\cdot,1)$ is a monoid,
  \emph{$0$ annihilates $\cdot$} (i.e., $0\cdot s_1=0 = s_1\cdot 0$)
  and \emph{$\cdot$ distributes over $+$} (i.e.,
  $(s_1+ s_2)\cdot s_3=s_1\cdot s_3+ s_2\cdot s_3$ and
  $s_3\cdot (s_1+ s_2)=s_3\cdot s_1+ s_3\cdot s_2$, for all
  $s_1,s_2,s_3\in S$).
\item A \emph{complete distributive lattice} is a partially ordered
  set $(L,\sqsubseteq)$ where for all subsets $L'\subseteq L$ of $L$
  the infimum $\bigsqcap L'$ and the supremum $\bigsqcup L'$
  w.r.t. the order $\sqsubseteq$ exists and infimum distributes over
  finite suprema:
  $(\ell_1\sqcup \ell_2)\sqcap \ell_3=(\ell_1\sqcap \ell_3)\sqcup
  (\ell_2\sqcap \ell_3)$ for all $\ell_1,\ell_2,\ell_3\in L$. Together
  with $\top=\bigsqcup L$ and $\bot=\bigsqcap L$, a complete
  distributive lattice forms a semiring
  $\mathbb L=(L,\sqcup,\sqcap,\bot,\top)$.
\item An $l$-monoid is a lattice $(L,\sqcup,\sqcap,\bot,\top)$
  together with a monoid $(L,\cdot,1)$ such that $\cdot$ distributes
  over $\sqcup$. If $\bot$ annihilates $\cdot$,
  i.e. $\bot\cdot\ell=\bot$ for all $\ell\in L$, then the $l$-monoid
  can be regarded as a semiring $(L,\sqcup,\cdot,\bot,1)$.
\end{itemize}

\subsection{Weighted Automata}

A weighted automaton (WA) can be understood as a non-deterministic
automaton that additionally carries weights from a given semiring
$\mathbb S$ on each transition, as well as a termination weight for
each state. Since the automata will be represented as matrices with
semiring entries in \paws, we will use a matrix notation throughout
the paper. For us, a matrix is a mapping of the form
$\alpha\colon X\times Y\to \mathbb{S}$, where $X$, $Y$ are index sets
indicating rows respectively columns. The set $X$ is typically the set
of states and in many cases we will simply choose $X = \{1,\dots,n\}$.

\begin{defi}
  Let $A$ be a finite set of alphabet symbols and $X$ be a set of
  states. Then a \emph{weighted automaton} is an
  $X\times(A\times X+\{\bullet\})$-matrix $\alpha$ with entries from a
  semiring $\mathbb{S}$. We write $x\xrightarrow{a,s}x'$ if
  $\alpha(x,(a,x'))=s$.
\end{defi}

For a weighted automaton $\alpha$, $\alpha(x,\bullet)$ denotes the
final weight of state $x\in X$ and $\alpha(x,(a,y))$ denotes the
weight of the $a$-transition from $x$ to $y$.

Rather than accepting or rejecting a word $w$, a state $x$
in a weighted automaton associates it with a value from $\mathbb
S$. This value can be obtained as follows: Take the sum of the weight
of all $w$-labelled paths $p$ starting in $x$. The weight of a path $p$
is the product of all transition weights along $p$, including the
termination weight. Consider for instance the following simple
weighted automaton over the field $\mathbb R$:
 
 \begin{center}
	\begin{tikzpicture}[x=2.5cm,y=1.5cm,double distance=2pt]
      \node[state] (a) at (1,2) {$A$} ;
      \node[state] (d) at (0,1) {$B$} ;
      \node[state] (e) at (2,1) {$C$} ;
      \node (terma) at (0.5,2) {} ;
      \node (termd) at (-0.5,1) {} ;
      \node (terme) at (2.5,1) {} ;
      \begin{scope}[->]
        \path[shorten <=1pt] 
          (a) edge node[above] {$a,2$} (d) 
          		edge node[above] {$a,3$} (e)
          		edge node[above] {$1$} (terma);
        \path 
          (d) edge[loop below] node {$b,2$} (d)
         			edge node[above] {$2$} (termd);
        \path 
          (e) edge[loop below] node {$b,2$} (e)
         			edge node[above] {$1$} (terme);
      \end{scope}
    \end{tikzpicture}
\end{center}

The weight of the word $ab$ in state $A$ can be computed as follows:
there are two paths to consider, $(A,a,B,b,B)$ and
$(A,a,C,b,C)$. The first path has the weight $2\cdot 2\cdot 2=8$ and
the second path has the weight $3\cdot 2\cdot 1=6$, so overall, the
weight of the word $ab$ in $A$ is $8+6=14$.

Formally, the language of a weighted automaton can be defined as
follows:

\begin{defi}[Language of a Weighted Automaton \cite{hwaDKV}]
  Let
  $\alpha\colon X\times (A\times X+\{\bullet\})\rightarrow\mathbb S$
  be a weighted automaton. The language
  $L_\alpha:A^*\rightarrow (X\to \mathbb{S})$ of $\alpha$ is recursively
  defined as
  \begin{itemize}
  \item $L_\alpha(\epsilon)(x) = \alpha(x,\bullet)$ where $\epsilon$
    is the empty word
  \item
    $L_\alpha(aw)(x) = \sum_{x'\in X} \alpha(x,(a,x'))\cdot L_\alpha(w)(x')
    $ for $a\in A$, $w\in A^*$
  \end{itemize}
  We will call $L_\alpha(w)(x)$ the weight that state $x$ assigns to
  the word $w$. Two states $x,y\in X$ are language equivalent if
  $L_\alpha(w)(x) = L_\alpha(w)(y)$ for all $w\in A^*$.
\end{defi}

For weighted automata, \paws offers two algorithms to decide which
pairs of states are language equivalent, i.e. assign the same weight
to all words. In general, this problem is undecidable, but for many
specific semirings (e.g. the reals) it is decidable.

The more generally applicable algorithm, called \fonta{Language
  Equ.(Complete)} in the tool, can be applied to any semiring and
yields, if it terminates, a complete characterisation of language
equivalence in the automaton. This algorithm is a generalised
partition refinement algorithm, based on the final chain construction
in coalgebra \cite{KK14}. The idea is to enumerate the weights that
each state assigns to all words, starting from the empty word
$\epsilon$. Whenever a word is found whose corresponding weight vector
is a linear combination of previously explored words, the vector can
be dropped and no words extending this word need to be explored. As
soon as all branches terminate, the algorithm terminates. Then, the
language equivalent states are exactly those which assign the same
weight to all words that were explored. It is worth noting that, in
contrast to traditional partition refinement algorithms, this
algorithm may require additional steps after the final partition of
the state set is already established. This is to be expected though,
because of the undecidability of language equivalence in general.

For complete and completely distributive lattices, and more generally,
completely distributive $l$-monoids, an optimisation up to congruence
is available (\fonta{Language Equ.(Up-To)}). This algorithm checks
whether two initial vectors are language equivalent by building a
language equivalence relation on the fly on vectors over the semiring
and pruning the relation by congruence closure, i.e. pairs of vectors
that are already in the congruence closure of previously found pairs
of vectors, are discarded \cite{bkk:up-to-weighted}. When compared to
\fonta{Language Equ.(Complete)}, additional optimisation occurs in the
form of pruning additionally vector pairs that are redundant because
of symmetry or transitivity. Moreover, the algorithm may perform much
better than \fonta{Language Equ.(Complete)} because only one pair of
vectors is being compared, instead of all pairs of states, so in
particular, a counterexample to language equivalence may be quickly
identified.

One semiring, where language equivalence is undecidable is the
tropical semiring over the natural numbers
$(\mathbb N_0\cup\{\infty\}, \mathsf{min}, +, \infty,0)$, as shown by
Krob \cite{Krob94theequality}. However, the universality problem is
decidable in this case. The universality problem asks, for any given
threshold $T\in\mathbb N_0$, whether the weight of all words from a given
initial vector is bounded by $T$.

Note that all algorithms for these problems require a method for
solving linear equations over the semiring.

\subsection{Conditional Transition Systems}

\paws also analyses a second type of systems: conditional transition
systems (CTS) \cite{bkks:cts-upgrades,ABHKMS12}. These systems are
related to featured transition systems
\cite{Classen:2013:FTS,DBLP:conf/icse/CordyCPSHL12} which can be used
for modelling software product lines. Different products are modelled
by a single transition system where transitions are annotated with the
products for which they are enabled. In addition CTS also have the
possibility to upgrade the product, thus enabling new transitions,
during runtime, i.e., they are adaptive.

More concretely, CTS are defined over a finite partial order of
conditions. Each transition is assigned to a downwards-closed set of
conditions under which the transition may be taken. Before execution,
one condition is fixed and all transitions that carry the respective
condition remain active and all other transitions remain
inactive. Afterwards, the CTS evolves just like a traditional labelled
transition system. At any point though, a change of conditions can
occur by going down in the order. If the condition is changed, so do
the active transitions and additional transitions may become
available. Note that due to the requirement that all transitions carry
a downwards-closed set of conditions, only additional transitions can
appear, no transition can be deactivated by performing a change in
conditions. Conditions can intuitively be considered as different
versions of a program, where a smaller condition signifies an improved
product to which a user may upgrade at any point. Using this
interpretation, CTS can be used to model and analyse such classes of
systems in particular for their behaviour when upgrades occur.

Also note the strong relation of partial orders to complete
distributive lattices via the Birkhoff duality
\cite{dp:lattices-order}: Every finite partial order gives rise to a
lattice by considering all downwards-closed sets of elements ordered
by inclusion. And vice versa, every finite distributive lattice can be
represented in this way. We will use this duality to obtain compact
representations of CTS.

\begin{defi}
  \label{CTS} Let $(\Phi,\leq)$ be a finite partially ordered
  set. We call the elements of $\Phi$ conditions.

  A \emph{conditional transition system} (CTS) is a triple $(X,A,f)$
  consisting of a set of states $X$, a finite set $A$ called the label
  alphabet and a function
  $f: X \times A \rightarrow (\Phi\rightarrow \mathcal{P}(X))$ mapping
  every ordered pair in $X \times A$ to a monotone function of type
  $(\Phi,\leq) \rightarrow (\mathcal{P}(X),\supseteq)$.

  As usual, we write $x\xrightarrow{a,\phi} y$ whenever
  $y\in f(x,a)(\phi)$.
\end{defi}

As indicated above, there is a dual representation, where a CTS
$(X,A,f)$ is represented as a matrix
$\beta\colon X\times (A\times X)\to \mathcal{O}(\Phi)$, where
$\mathcal{O}(\Phi)$ is the set of all downward-closed subsets of
$\Phi$ (wrt.\ $\le$). We define
$\beta(x,(a,y)) = \{\phi\in \Phi\mid y\in f(x,a)(\phi)$, which is --
due to monotonicity -- always a downward-closed set. Furthermore
$\mathcal{O}(\Phi)$ is a lattice, ordered by inclusion, with
operations union and intersection, hence a semiring. However, here we
are interested in bisimilarity rather than language
equivalence. Furthermore the lattice elements will play the role of
guards rather than weights.

\medskip

Now consider the following CTS defined over the set of
conditions $\{\phi,\phi'\}$, where $\phi'\leq\phi$.

\begin{center}
  \begin{tikzpicture}[x=2.5cm,y=1.5cm,double distance=2pt]
    \node[state] (a) at (1,2) {$A$} ;
    \node[state] (d) at (0,1) {$B$} ;
    \node[state] (e) at (2,1) {$C$} ;
    \begin{scope}[->]
      \path[shorten <=1pt] 
      (a) edge node[left] {$a,\{\phi'\}$} (d) 
      edge node[right] {$a,\{\phi,\phi'\}$} (e);
      \path 
      (d) edge[loop below] node {$b,\{\phi,\phi'\}$} (d);
      \path 
      (e) edge[loop below] node {$b,\{\phi'\}$} (e);
    \end{scope}
  \end{tikzpicture}
\end{center}

Assume that we are starting in state $A$ under the initial condition
$\phi$. Then only the transitions from $B$ to $B$ and from $A$ to $C$
are available. If we choose to make a step from $A$ to $C$ via the
$a$-labelled edge, we cannot do any further steps, unless we first
perform an upgrade to $\phi'$, which allows us to use the $b$-loop in
state $C$. Alternatively, starting in $A$, we could also perform an
upgrade right away and gain the option to take an $a$ transition to
$B$ instead of transitioning to $C$.

For CTS we are interested in conditional bisimulation. Two states are
conditionally bisimilar, if there exists a conditional bisimulation
relating the states. A conditional bisimulation is a family of
traditional bisimulations $R_\phi$, one for each condition $\phi$, on
the respective underlying transition systems. For two conditions
$\phi'\leq\phi$ it must hold that $R_{\phi'}\supseteq R_\phi$, which
intuitively means that if two states are bisimilar under $\phi$, they
must also be bisimilar under every smaller condition $\phi'$.
Furthermore, the standard transfer property for bisimulations must be
satisfied.

In \cite{bkks:cts-upgrades} we have shown how to model a small
adaptive routing protocol as CTS.

Bisimulation for CTS is computed with a modified partition refinement
algorithm akin to the partition refinement algorithm to compute
bisimilarity for labelled transition systems. This algorithm uses the
duality between finite partially ordered sets and finite distributive
lattices to compute and propagate in parallel the partition refinement
for all conditions at once. The BDD based-implementation uses the same
basic algorithm, but represents lattice elements via BDDs, which can
often lead to a very compact representation, which in turn makes the
algorithm viable for significantly larger sets of conditions.


\smallskip

Summarizing, the problems \paws solves and the corresponding
algorithms and semirings are displayed in the following table:

\begin{center}\scalebox{1}{
  \begin{tabular}{|c|c|c|c|}\hline
    Problem & Algorithm & Semiring & Model \\ \hline
    Language equivalence (all pairs) & \fonta{Language Equ.(Complete)} & 
    any semiring & WA \\
    Language equivalence (initial vectors) & \fonta{Language Equ.(Up-To)} &
    $l$-monoids, lattices & WA \\
    Universality Problem & \fonta{Universality} & tropical Semiring & WA \\
    Conditional bisimilarity & \fonta{CTS Bisimilarity} & finite lattice & 
    CTS \\ \hline
\end{tabular}}
\end{center}

\section{Design and Usage}

In this section, we give an overview of some design decisions and the
usage of the tool. First, we explain the basic structure and then
discuss some of the problems and math-related features of the
tool. Furthermore we explain how to work with \paws.

\subsection{Design}

\paws is a Windows tool offering a complete graphical interface,
developed in Microsoft's Visual Studio using C\#. The program is
divided into two autonomous components:

\begin{figure}[h]
  \centering
  \includegraphics[width=0.6\textwidth]{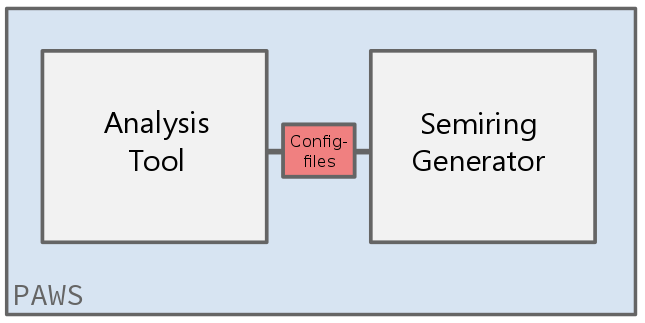}
  \label{fig_overview}
\end{figure}

Both components are designed according to the MVC (Model View
Controller) pattern. In the sequel, we will discern these two program
parts, as their interaction is rather limited, allowing them to be
considered as separate concerns.

\subsubsection{The Semiring Generator}

The development of the semiring generator started in a master's thesis
\cite{m:gen-werkzeug-sprachaequ}. It supports five different
generation processes, which, for clarity, are equipped with five
separate input forms. The semiring generator supports three fully
automated cases:

\begin{itemize}
\item Direct products
\item Fields of fractions
\item Field extensions for $\mathbb{Q}$
\end{itemize}

Furthermore, there are two options to generate code based on user
implementations:

\begin{itemize}
\item $l$-monoids
\item Arbitrary semirings 
\end{itemize}

Note that generation of finite lattices and modulo rings is less
involved and is done directly within the analysis component. 

For the generation of source code we use \textsl{CodeDOM} of the
\textsl{.NET} Framework, which enables code generation based on object
graphs. The five processes are implemented via one superclass and
three derived classes. The superclass contains all methods for
creating if-statements, for-loops or useful combinations of these,
based on predefined patterns.  Except for the direct product
generator, every class uses the constructor generating methods of the
superclass. Most of the differences between the classes are reflected
in the methods for generating the binary operators. Concerning methods
for solving linear equations we are offering two templates: a
straightforward implementation of the Gaussian algorithm and one
method for $l$-monoids, based on the residuum operator
\cite{opac-b1085541}.

With code generation, one always has to face the question of how to
give the program, specifically the analysis component of \paws, access
to the newly generated classes.
%
%
We decided to use the \textsl{Microsoft.Build.Execution} namespace for
updating the analysis tool. This decision avoids creating multiple DLL
files, which would be the case with a pure reflection-based solution
to the problem. However, reflection is used to solve another
difficulty. Due to the combination of different semirings or data
types as elements of a new semiring, a dynamic approach is required to
enable the automated generation of constructors with a string
parameter, specifying the semiring value.

\paws also manages the names of the semiring classes in individual
text files. First, this prevents a user from overwriting an already
fixed class name. Another advantage is that by using
\textsl{System.Activator}, an instance can be created dynamically
during runtime without knowing the class name at the level of program
design. Hence, both components use consistent config files, which is
ensured by updating the corresponding config file if one of the
program parts creates or deletes a semiring. But when deleting a
semiring that has already been used to create and store an automaton
or transition system, conflicts may occur with the serialised objects
and thus with the user's storage files. Therefore, deletion of
semirings must be dealt with separately. A semiring can only be
deleted or modified if it has not been used in the previous session
within an automaton.

 
\subsubsection{The Analysis Tool}

As already mentioned, for both program parts MVC is used to implement
the user interface. The main focus of \paws is to give the user the
tools to define an automaton in order to be able to subsequently
analyse it with the supported algorithms. In order to implement this
as dynamically as possible, we have opted for a generic
implementation. Therefore, the class
$\mathit{Matrix}\langle T\rangle$ (where $T$ is the generic type of
the semiring), which implements automata in a matrix representation,
is the core of the tool's architecture.

\begin{figure}[h]
  \centering
  \includegraphics[width=0.4\textwidth]{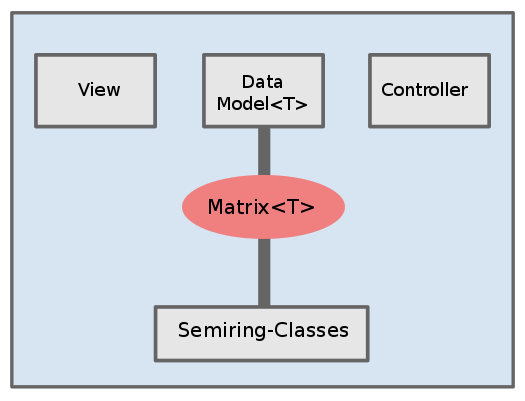}
  \label{fig_aws}
\end{figure}

The main argument for the generic approach is that the algorithms for
the analysis of an automaton are based on the basic operators of a
semiring. A further method is needed to solve systems of linear
equations, which is also defined for each semiring within \paws. It is
therefore recommended to use a generic class that includes all the
methods of analysis, which in turn dynamically call the corresponding
operators or methods of the currently used semiring. This dynamic
approach is thus combined with reflection.

Consequently, the management of automata created by the user requires a
generic implementation as well. The model for the matrices is
therefore also generic and thus it makes sense to manage several
semiring models by the controller.

Because of the generic approach, automatically checking the
correctness of the user input proves to be problematic when the user
has generated his or her own semiring and has implemented a string
constructor to read semiring elements from input strings without input
verification. In this case \paws can not check whether the input is
well-formed, this has to be taken care of in the user-implemented
method. Such semirings can not be further used to generate other
semirings.


A further design decision is that one can create finite semirings
directly within the analysis component of \paws. The reason why we
chose to not move this option to the generator is that for finite
lattices and modulo rings over the integers no new classes have to
be generated nor is there any need for new source code at all. In this
case, it is sufficient to configure template classes for the
corresponding semirings via static variables that contain the required
information about the semiring. The operators are designed to behave
according to the configuration of the class. In such cases, the
analysis program must also access the configuration files in order to
make the extensions known to the semiring generator.

As an additional feature, we have also integrated
GraphViz\footnote{\texttt{www.graphviz.org}} into the tool, as it
allows visualization of weighted automata, if desired by the user. 

\subsection{Usage}

We will discuss the usage of the tool separately for the two
individual components of \paws, hence this section contains
subsections giving details about the following two components.

\begin{itemize}
\item The \emph{semiring generator} to build and provide the required
  semirings over which automata can be defined. This generator is used
  to generate semirings that cannot be obtained in a fully automated
  way and supports some form of automatic generation.
\item The \emph{analysis tool} that allows the user to choose a
  previously generated semiring, one of the semirings that come
  built-in with \paws or to build a lattice or modulo ring, and then
  to define automata in a matrix representation over the chosen
  semiring. Matrices are then interpreted as weighted automata (WA) or
  conditional transition systems (CTS) and can be used to compute
  language equivalence for weighted automata with two different
  approaches, decide the threshold problem for weighted automata over
  the tropical semiring of natural numbers or to compute the greatest
  bisimilarity of a CTS.
\end{itemize}

\subsubsection{The Semiring Generator}

The semiring generator is used to generate the semirings under
consideration. In order for a semiring to be usable within the context
of \paws, the structure needs to define the following components:

\begin{itemize}
\item A universe which contains all elements of the semiring. All
  predefined datatypes from C\#, as well as combinations of them can
  serve as universes.
\item An addition operator $+$ of the semiring.
\item A multiplication operator $\cdot$ of the semiring.
\item One() and Zero() methods, which return the units of addition and multiplication.
\item A method for solving linear equations over the semiring.
\end{itemize}

\begin{figure}[h]
  \centering
  \includegraphics[width=0.9\textwidth]{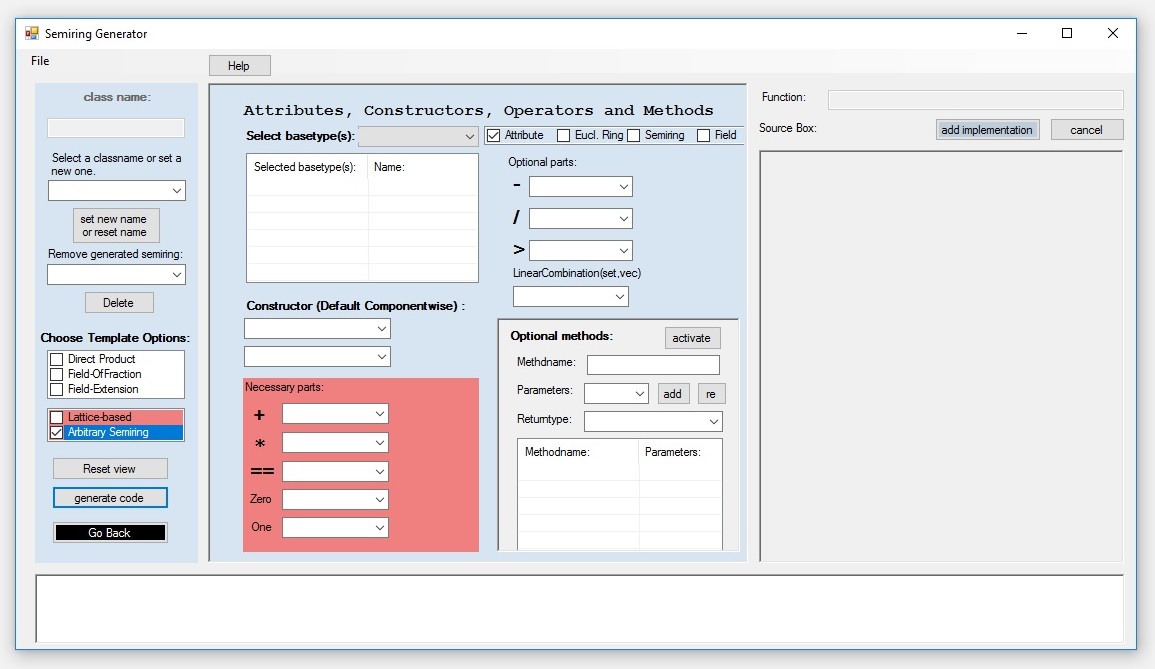}
  \label{PAWS_02}
\end{figure}

While the first four components are necessary to define a semiring in
a mathematical context either way, the procedure to solve linear
equations is an additional requirement -- one that \paws aims at
reducing as much as possible -- but in order to provide the greatest
amount of flexibility possible, the tool offers the option to define a
procedure to solve linear equations from scratch.

\subsubsection{The Analysis Tool}

The analysis component is designed to offer generic algorithms
applicable to numerous predefined or user-defined semirings. Some of
the algorithms can however only be used with specific (types of)
semirings. The most general algorithm is the language equivalence
check, for which all semirings are eligible. Conditional transition
systems are only defined over lattices, therefore, the bisimulation
check is limited to lattice structures. However, the user still has
the choice between two different ways of dealing with lattices:
representing elements of the lattice as downwards-closed sets of
irreducibles via the Birkhoff duality \cite{dp:lattices-order},
applicable to all finite distributive lattices, or representing them
using binary decision diagrams (BDDs). The BDD variant is more
restrictive and mainly designed for the application to CTS.  Here, the
irreducibles are required to be full conjunctions of features from a
base set of features, ordered by the presence of distinct upgrade
features. Lastly, the threshold check can only be performed over a
single semiring, the tropical semiring over natural numbers.

\begin{figure}[h]
  \centering
  \includegraphics[width=0.9\textwidth]{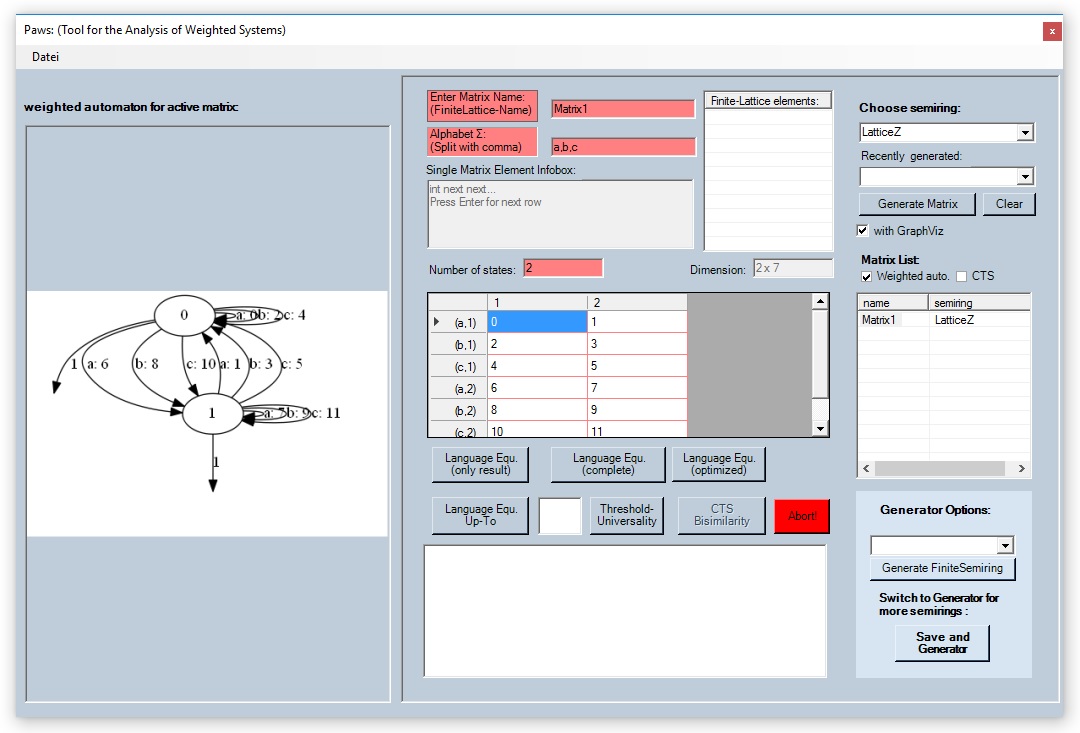}
  \label{PAWS_01}
\end{figure}

\medskip

The general workflow of the analysis tool is as follows:

\begin{itemize}
\item[$\triangleright$] Choose a semiring
\item[$\triangleright$] Generate a matrix over this semiring,
  representing a weighted automaton or a conditional transition system

  \emph{Alternatively:} Choose the matrix from a list of matrices that
  have been generated previously
\item[$\triangleright$] Start the algorithm and provide -- if necessary -- additional
  input
\end{itemize}

Additional input comes in two forms: initial vectors and the threshold
to be checked against in case of the threshold algorithm. Depending on
the semiring of choice, questions regarding language equivalence might
not be decidable, leading to non-termination of the corresponding
procedure in \paws. In order to deal with this problem and to allow
abortion of an overlong computation, the actual computation is
delegated to a separate thread that can at any time be aborted. In
that case all intermediate results are discarded.
 
Note that only the two language equivalence-based algorithms can run
into non-termination issues. For the CTS bisimulation check, as well
as the threshold problem on the tropical semiring of natural numbers,
termination is always guaranteed. However, the runtime of CTS
bisimulation check can be doubly exponential in the number of features
under consideration -- because the lattice is the set of all possible
configurations, which in turn are all possible conjunctions over the
features. Using the BDD-based implementation of lattices -- which is
particularly suited to the needs of CTS modelling -- this explosion is
mitigated in many cases, but it can not be ruled out completely. On
the other hand, the BDD-based modelling only allows for special
lattices to be modelled, i.e. those that arise as the lattices
constructed from a set of features and upgrade features, whereas the
variant called \textsl{FiniteLattice} allows for arbitrary (finite,
distributive) lattices to be represented. In this case, lattices are
represented via the partial order of irreducible elements, using
Birkhoff's representation theorem \cite{dp:lattices-order}.

\section{GUI Overview of \paws}

In this section, we give a short overview of the GUI. First, we will
present the semiring generator and how to start a semiring generation
process. We then illustrate the various possibilities to use the
analysis component of \paws in a short overview.


\subsection{The Semiring Generator}

\begin{figure}[h]
  \begin{minipage}[t]{0.475\textwidth}
    \centering
    \includegraphics[height=45mm]{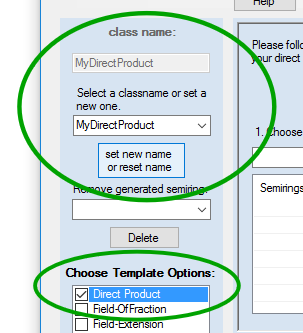}
    \caption{Lower ellipse: Choose an input mask. Top ellipse: Specify
      the name of the new semiring class.}
    \label{fig_dp_01}
  \end{minipage}
  \hfill
  \begin{minipage}[t]{0.475\textwidth}
    \centering
    \includegraphics[height=45mm]{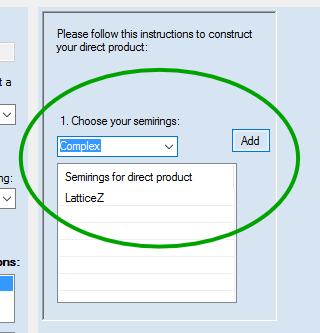}
    \caption{Specify a direct product, add semirings as member
      fields to the class.}
    \label{fig_dp_02}
  \end{minipage}
\end{figure}

\begin{figure}[!h]
  \centering
  \includegraphics[width=0.7\textwidth]{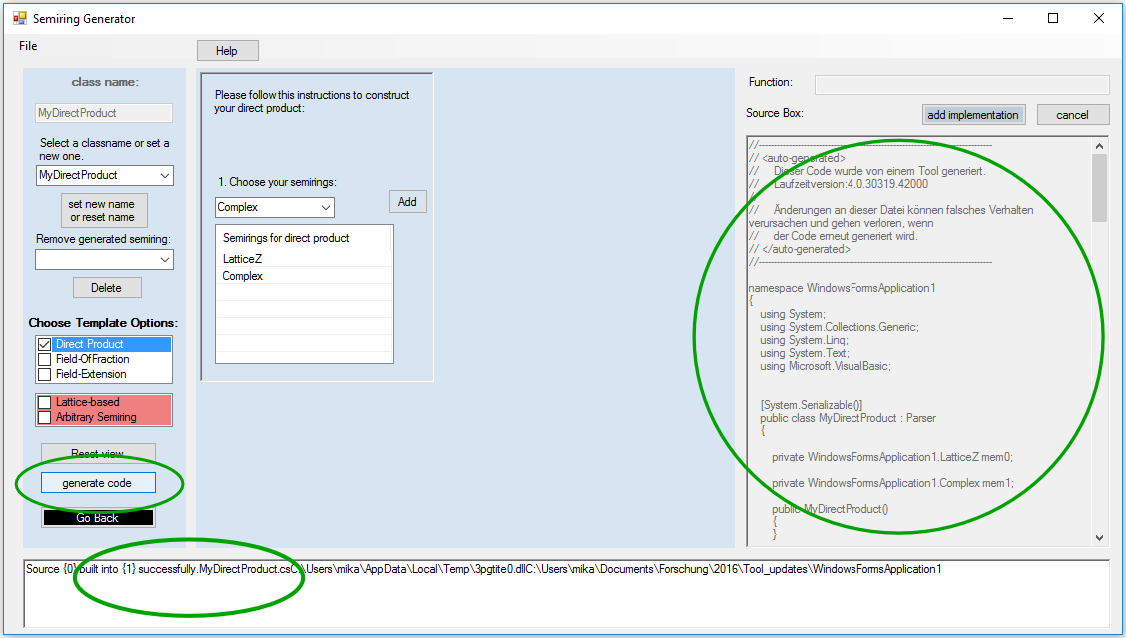}
  \caption{Pressing the circled button will generate the code visible in the
    right-hand text-box. The console at the bottom informs the user,
    whether the source code was compiled and successfully added to the
    analysis component of \paws.}
  \label{fig_dp_03}
\end{figure}

First, the user hast to choose one of the offered generation
templates, for example the direct product input mask
(Figure~\ref{fig_dp_01} and~\ref{fig_dp_02}).  Then, in
Figure~\ref{fig_dp_03} the console informs the user that the generated
source code is compilable. In case the user specifies some code on his
or her own, the console will display suitable compilation error
messages.



\subsection{The Analysis Tool}

In this section the use of the \paws analysis component is presented
in a brief overview. The illustrations serve to explain our intuition
behind the design and the use of \paws.

\begin{figure}[h]
  \begin{minipage}[t]{0.475\textwidth}
    \centering
    \includegraphics[height=45mm]{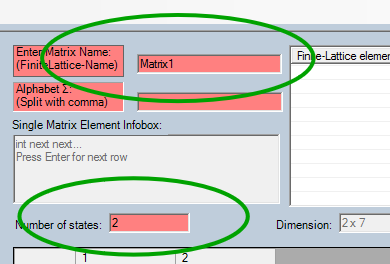}
    \caption{Top ellipse: First determine the name of the
      automaton. Bottom ellipse: Type in the number of states.}
    \label{fig_aw_01}
  \end{minipage}
  \hfill
  \begin{minipage}[t]{0.475\textwidth}
    \centering
    \includegraphics[height=45mm]{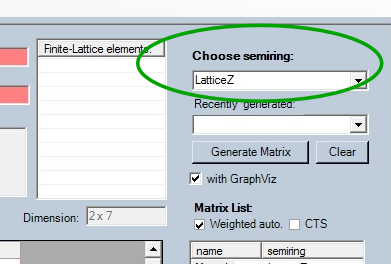}
    \caption{Choose the semiring.}
    \label{fig_aw_02}
  \end{minipage}
\end{figure}

\begin{figure}[h]
  \centering
  \includegraphics[height=40mm]{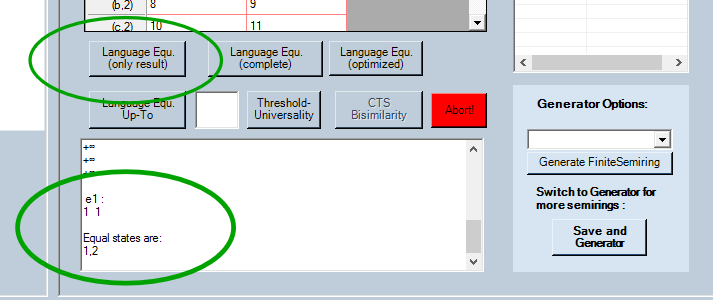}
  \caption{After starting one of the supported algorithms for a
    transition system, the analysis result will be displayed in the
    bottom text area.}
  \label{fig_aw_05}
\end{figure}

In Figures~\ref{fig_aw_01} and~\ref{fig_aw_02} the first steps for
creating a weighted automaton are illustrated. After generating a
matrix, the user can
choose one of the available algorithms and wait until the result of
the computation is displayed inside the text area (Figure
\ref{fig_aw_05}).


With the \paws analysis component, besides automata, also finite
semirings can be generated and stored for further semiring generation
as well as for the analysis.

\subsection{Evaluation}

Here we describe several case studies and list runtime results.
Before we discuss the results, let us first consider the theoretical
complexity results, listed in Table~\ref{fig:overview problems}, where
we consider various semirings (in the left column) and various
problems (in the top row). 

\begin{table}[h]
  \centering
  \begin{tabular}{l|ccc}
    & equivalence & inclusion & threshold \\ \hline
    \\
    real & $\mathsf{P}$ & undecidable & undecidable \\
    numbers & {\footnotesize \cite{t:polynomial-equ-prob-automata}} &
    \multicolumn{2}{c}{\footnotesize
      \!\!\!\!\!\cite{p:prob-automata}} \\ \\
    tropical & undecidable & undecidable & $\textsf{PSPACE}$-cmpl.
    \\
    semiring & \multicolumn{2}{c}{\footnotesize \cite{Krob94theequality}} &
    \hspace{-1cm}{\footnotesize \cite{abk:decidable-weighted-automata}} \\
    \\
    distr. & $\textsf{PSPACE}$-cmpl. & $\textsf{PSPACE}$-cmpl.
    & $\textsf{PSPACE}$-cmpl. \\
    lattices & \multicolumn{3}{c}{\footnotesize
      \cite{doi:10.1142/S0129054110007192}} \\
  \end{tabular}
  \caption{Overview of the corresponding complexity classes for
    considered weighted automata problems.}
  \label{fig:overview problems}
\end{table}

The main focus of \paws\ is on weighted automata and language
equivalence, hence we will conduct our experiments for this
problem. For extensive experiments concerning the universality problem
see \cite{bkk:up-to-weighted}. Experiments on conditional transition
systems were conducted in \cite{bkks:cts-upgrades}.

We will consider the decidable cases, namely the reals, as well as
another (finite) ring, and a distributive lattice. More concretely, we
will evaluate the algorithm \fonta{Language Equ.(Complete)} on three
different semirings: the reals $(\mathbb{R},+,\cdot,0,1)$, the finite ring
$(\mathbb{Z}_{100},+_{100},\cdot_{100},0,1)$ (where addition and
multiplication is modulo $100$), and $\mathit{LatticeZ}$, the lattice
$(\mathbb{Z},\min,\max,\infty,-\infty)$ restricted to the interval
$[-1000,1000]$. Note that \fonta{Language Equ.(Complete)} determines
for all pairs of states whether they are equivalent.

\paws was executed on an Intel Core 2 Quad CPU Q9550 at 2.83 GHz with
4 GB RAM, running Windows 10. Our test cases were constructed as
follows: we considered different numbers of states ($|X|$) and also
varied the probability of the existence of a transition between a pair
of states ($p_\mathit{Tr}$). Usually we set $p_\mathit{Tr} = 0.5$, but
we will also compare to other values. Weights are set randomly and the
default alphabet size was~$2$.

For each semiring and each chosen combination of parameters we
randomly generated 1000 weighted automata and observed the runtime
behaviour. We considered various percentiles, for instance the $50\%$
and $90\%$ percentiles: the $50\%$ percentile is the median and the
$90\%$ percentile means that $90\%$ of the runs were faster and $10\%$
slower than the time given in the respective field. Analogously for
the other percentiles.

As expected, the runtimes are heavily dependent on the semiring. This
is due to two reasons: first, the algorithms \fonta{Language
  Equ.(Complete)} internally uses procedures for solving linear
equations and the complexity of those procedures, which depend on the
semiring, plays a major role in the runtime. Second, as explained in
Section~\ref{sec:preliminaries}, the algorithm lists words of
increasing length and associates them with weight vectors. A vector
need not be considered if it is a linear combination of existing
vectors. Hence, if the corresponding semimodule has few generators --
as it is for instance the case for a vector space of limited dimension
over a field -- the search space can be cut off quickly.

We first consider the two rings: the reals and $\mathbb{Z}_{100}$ (see
Table~\ref{fig:experiments}). For the reals, the Gaussian algorithm
has a polynomial runtime $\mathcal O(|X|^3)$ in the number of states.
Due to the fact that the search can be cut off after at most $|X|$
vectors, there are only minor variations in the runtime over the
various percentiles. Furthermore the alphabet size and the probability
$p_\mathit{Tr}$ have little impact on the runtime.  Instead the times
are mainly dependent on the number of states.


\begin{table}[h]
  \centering
  {\footnotesize
    \begin{tabular}{|r|r|r|r|r|}
      \hline
      $(|X|,p_\mathit{Tr})$&Semiring&$50\%$&$90\%$&$95\%$\\
      \hline
      (10,0.5)&$\mathbb{R} $&28&29&30\\&$\mathbb{Z}_{100}$&42&73&2307\\\hline
      (15,0.5)&$\mathbb{R} $&98&100&101\\&$\mathbb{Z}_{100}$&252&263&267\\\hline
      (20,0.5)&$\mathbb{R} $&359&361&362\\&$\mathbb{Z}_{100}$&518&523&528\\\hline
      (25,0.5)&$\mathbb{R} $&712&717&719\\&$\mathbb{Z}_{100}$&1034&1058&1081\\\hline
      (30,0.5)&$\mathbb{R} $&1327&1333&1336\\&$\mathbb{Z}_{100}$&1983&3509&3548\\\hline
      (35,0.5)&$\mathbb{R} $&3648&3660&3664\\&$\mathbb{Z}_{100}$&5242&5290&5304\\\hline
      (40,0.5)&$\mathbb{R} $&5183&5197&5202\\&$\mathbb{Z}_{100}$&7654&7924&7960\\\hline
      (45,0.5)&$\mathbb{R} $&7400&7419&7425\\&$\mathbb{Z}_{100}$&11492&11538&11551\\\hline
      (50,0.5)&$\mathbb{R} $&10471&10509&10720\\&$\mathbb{Z}_{100}$&15673&15720&15738\\\hline
    \end{tabular}
  }\caption{Runtime results in milliseconds on randomly generated
    automata over two different semirings}
  \label{fig:experiments}
\end{table}

There are different effects for the ring $\mathbb{Z}_{100}$, which
also uses the Gauss method, but the runtime is particularly influenced
by the subsequent application of the Hensel lifting, which depends on
the one hand on the prime factor decomposition
$q = \prod_{i=0}^{k} p_i^{e_i}$, where $k$ is the number of prime
factors (here: $q=100$), and second on the number of zero lines that
occur in computation of the row echelon form. It is not previously
known which combination of the free variables leads to a solution
\cite{Hensel:lifting,m:gen-werkzeug-sprachaequ}. The cases in which all those possibilities have to be
enumerated are very rare, however, these cases do occur and they may
cause a timeout after 5~hours. In order to see this effect, consider
the $99.9\%$ percentiles in Table~\ref{fig:hensel_worstcase}. An
additional, very interesting, observation, which we have made while
testing the implementation, is that the worst case is less frequent in
practice with an increasing number of states (see
Table~\ref{fig:hensel_worstcase}). A possible explanation for this
could be the decreasing probability for the occurrence of zero lines
with an increase in the number of states.

The algorithm is much less efficient in the case of the lattice of
integers $\mathit{LatticeZ}$ (see Table~\ref{fig:lattice}), although
the procedure for solving a single solving equation system, based on
the residuum operation \cite{opac-b1085541}, has linear
runtime. However, this does not help, since we can not guarantee to
find a small set of generators in the semimodule and hence there is no
limit on the number of vectors which are enumerated. Already for
$|X| = 15$ states it is infeasible to obtain results. The situation
becomes worse when the alphabet size increases, since the number of
vectors will also grow. This is different for the case of the reals
and $\mathbb{Z}_{100}$, since there the number of states limits the
number of columns.


\begin{table}[h]
  \centering
  \begin{tabular}{|r|r|r|r|r|r|r|r|}
    \hline
    $(|X|,p_\mathit{Tr})$&Semiring&$50\%$&$85\%$&$90\%$&$95\%$&$99\%$&$99.9\%$\\
    \hline
    (10,0.5)&$\mathbb{Z}_{100}$&42&68&73&2307&7041&time-out\\\hline
    (15,0.5)&$\mathbb{Z}_{100}$&252&258&263&267&897&1787758\\\hline
    (20,0.5)&$\mathbb{Z}_{100}$&518&521&523&528&532&time-out\\\hline
    (25,0.5)&$\mathbb{Z}_{100}$&1034&1044&1058&1081&1821&time-out\\\hline
    (30,0.5)&$\mathbb{Z}_{100}$&1983&3493&3509&3548&3745&3881\\\hline
    (35,0.5)&$\mathbb{Z}_{100}$&5242&5283&5290&5304&5422&5501\\\hline
  \end{tabular}
  \caption{Runtimes in milliseconds of \fonta{Language Equ.(Complete)}
    on randomly generated automata over $\mathbb{Z}_{100}$  .}
  \label{fig:hensel_worstcase}
\end{table}

\begin{table}[h]
  \centering	
\begin{tabular}{|r|r|r|r|r|r|r|}
	\hline
	$(|X|,p_\mathit{Tr})$&Semiring&$50\%$&$85\%$&$90\%$&$95\%$&$99\%$\\
	\hline
(5,0.5)&$LatticeZ$&41&242&347&758&3186\\\hline
(10,0.5)&$LatticeZ$&18808&287920&564525&1975251&16321319\\\hline
\end{tabular}
  \caption{Runtimes in milliseconds of \fonta{Language Equ.(Complete)}
    on randomly generated automata over $\mathit{LatticeZ}$.}
  \label{fig:lattice}
\end{table}

Note, that while the performance of algorithm \fonta{Language
  Equ.(Complete)} is bad in the case of $\mathit{LatticeZ}$, the
on-the-fly up-to technique \fonta{Language Equ.(Up-To)} performs much
better. At first sight, such a comparison seems unfair, since
\fonta{Language Equ.(Up-To)} tests language equivalence only for a
single pair of states, whereas the other method considers all pairs of
states at once. However, the advantage is significant enough that the
up-to technique can be expected to be more performant even when it is
applied to all pairs of states, one after the other. In fact, in the
case of randomly generated automata with ten states, there are 90
pairs of states to consider and the running time we have recorded for
\fonta{Language Equ.(Up-To)} is lower than 38 milliseconds in 99\%
of all cases. Thus, even if we take the value of the 99\% percentile
for all pairs of states, we end up with $90\cdot 38$ milliseconds,
which is significantly lower than the median runtime for
\fonta{Language Equ.(Complete)}.

For randomly generated automata, it is rare to find two
language-equivalent states, in fact, only very few steps are typically
required to find a word that separates randomly generated
automata. While \fonta{Language Equ.(Up-To)} may stop its computation
after finding a single word separating the initial pair of states, it
is not sufficient for termination of \fonta{Language Equ.(Complete)}
to find witnesses of absence of language equivalence. The intermediate
results of \fonta{Language Equ.(Complete)} also need to span the same
semimodule for termination of the algorithm, which may take numerous extra steps after all non-equivalent states have been separated.

\section{Conclusion, Future Work and Related Work}

We have seen that \paws is a flexible tool to analyse the behaviour of
weighted automata and conditional transition systems. The generic
approach allows for adding new semirings with a varying degree of
support by the tool itself. 

Concerning related approaches, we are not aware of analysis tools for
language equivalence and the threshold problem for weighted automata.

For the problem of generating semirings dynamically, there exists
previous work for solving fixpoint equations over semirings by
Esparza, Luttenberger and Schlund \cite{Esparza2014}. They introduce
\textsc{FPsolve}, a $\verb!C++!$ template based tool for solving
fixpoint equations over semirings. That is, the tool has a different
application scenario than ours. However, the tools share similarities
since in \textsc{FPsolve} the user also has the possibility to
generate new semirings. For this, only the addition, multiplication
and Kleene star must be implemented. However, a string constructor
must also be specified without automatic support and the main method
must be adjusted with the corresponding command-line. \paws is
designed to enable the generation of new semirings for solving linear
equations using a graphical user interface and does not change already
existing code, which is not part of a semiring class.

In contrast to this, work has already been done on an analysis tool
for featured transition systems -- which are basically CTS without a
notion of upgrades -- to analyse software product lines
wrt. simulation. In their work \cite{DBLP:conf/icse/CordyCPSHL12},
Cordy et al.\ have implemented their model using BDDs as well, yielding
a similar speed up as our own approach. The significant differences
here lie in the notion of behaviour, since Cordy et al. have focused
on simulation relations, whereas we are concerned with bisimulations.
Furthermore we capture a notion of upgrade and thus support partial
orders of products instead of just abritrary sets of products.

We intend to develop \paws further in several ways. We are looking for
new classes of semirings where solutions of linear equations can be
effectively computed, in order to equip those semirings with an
improved support from \paws. Furthermore, we are interested in
analysing more extensive case studies, where we will use \paws to
conduct all required analyses.

\bibliographystyle{eptcs}
\bibliography{references_doi}

\begin{thebibliography}{10}
\providecommand{\bibitemdeclare}[2]{}
\providecommand{\surnamestart}{}
\providecommand{\surnameend}{}
\providecommand{\urlprefix}{Available at }
\providecommand{\url}[1]{\texttt{#1}}
\providecommand{\href}[2]{\texttt{#2}}
\providecommand{\urlalt}[2]{\href{#1}{#2}}
\providecommand{\doi}[1]{doi:\urlalt{http://dx.doi.org/#1}{#1}}
\providecommand{\bibinfo}[2]{#2}

\bibitemdeclare{inproceedings}{ABHKMS12}
\bibitem{ABHKMS12}
\bibinfo{author}{Ji{\v{r}}\'i \surnamestart Ad\'amek\surnameend},
  \bibinfo{author}{Filippo \surnamestart Bonchi\surnameend},
  \bibinfo{author}{Mathias \surnamestart H\"{u}lsbusch\surnameend},
  \bibinfo{author}{Barbara \surnamestart K\"{o}nig\surnameend},
  \bibinfo{author}{Stefan \surnamestart Milius\surnameend} \&
  \bibinfo{author}{Alexandra \surnamestart Silva\surnameend}
  (\bibinfo{year}{2012}): \emph{\bibinfo{title}{A Coalgebraic Perspective on
  Minimization and Determinization}}.
\newblock In: {\sl \bibinfo{booktitle}{Proc. of FOSSACS '12}},
  \bibinfo{publisher}{Springer}, pp. \bibinfo{pages}{58--73},
  \doi{10.1007/978-3-642-28729-9\_4}.
\newblock \bibinfo{note}{{LNCS/ARCoSS} 7213}.

\bibitemdeclare{inproceedings}{abk:decidable-weighted-automata}
\bibitem{abk:decidable-weighted-automata}
\bibinfo{author}{Shaull \surnamestart Almagor\surnameend}, \bibinfo{author}{Udi
  \surnamestart Boker\surnameend} \& \bibinfo{author}{Orna \surnamestart
  Kupferman\surnameend} (\bibinfo{year}{2011}): \emph{\bibinfo{title}{What's
  Decidable about Weighted Automata?}}
\newblock In: {\sl \bibinfo{booktitle}{Proc. of ATVA '11}},
  \bibinfo{publisher}{Springer}, pp. \bibinfo{pages}{482--491},
  \doi{10.1007/978-3-642-24372-1\_37}.
\newblock \bibinfo{note}{{LNCS} 6996}.

\bibitemdeclare{inproceedings}{bls:conjugacy}
\bibitem{bls:conjugacy}
\bibinfo{author}{Mariel-Pierre \surnamestart B\'{e}al\surnameend},
  \bibinfo{author}{Slyvain \surnamestart Lombardy\surnameend} \&
  \bibinfo{author}{Jacques \surnamestart Sakarovitch\surnameend}
  (\bibinfo{year}{2006}): \emph{\bibinfo{title}{Conjugacy and Equivalence of
  Weighted Automata and Functional Transducers}}.
\newblock In: {\sl \bibinfo{booktitle}{Prof. of CSR '06}},
  \bibinfo{publisher}{Springer}, pp. \bibinfo{pages}{58--69},
  \doi{10.1007/11753728\_9}.
\newblock \bibinfo{note}{{LNCS} 3967}.

\bibitemdeclare{inproceedings}{bkks:cts-upgrades}
\bibitem{bkks:cts-upgrades}
\bibinfo{author}{Harsh \surnamestart Beohar\surnameend},
  \bibinfo{author}{Barbara \surnamestart K\"onig\surnameend},
  \bibinfo{author}{Sebastian \surnamestart K\"upper\surnameend} \&
  \bibinfo{author}{Alexandra \surnamestart Silva\surnameend}
  (\bibinfo{year}{2017}): \emph{\bibinfo{title}{Conditional Transition Systems
  with Upgrades}}.
\newblock In: {\sl \bibinfo{booktitle}{Proc. of TASE '17 (Theoretical Aspects
  of Software Engineering)}}.
\newblock \bibinfo{note}{To appear}.

\bibitemdeclare{inproceedings}{bkk:up-to-weighted}
\bibitem{bkk:up-to-weighted}
\bibinfo{author}{Filippo \surnamestart Bonchi\surnameend},
  \bibinfo{author}{Barbara \surnamestart K\"onig\surnameend} \&
  \bibinfo{author}{Sebastian \surnamestart K\"upper\surnameend}
  (\bibinfo{year}{2017}): \emph{\bibinfo{title}{Up-To Techniques for Weighted
  Systems}}.
\newblock In: {\sl \bibinfo{booktitle}{Proc. of TACAS '17, Part~I}},
  \bibinfo{publisher}{Springer}, pp. \bibinfo{pages}{535--552},
  \doi{10.1007/978-3-662-54577-5\_31}.
\newblock \bibinfo{note}{{LNCS} 10205}.

\bibitemdeclare{inproceedings}{bp:checking-nfa-equiv}
\bibitem{bp:checking-nfa-equiv}
\bibinfo{author}{Filippo \surnamestart Bonchi\surnameend} \&
  \bibinfo{author}{Damien \surnamestart Pous\surnameend}
  (\bibinfo{year}{2013}): \emph{\bibinfo{title}{Checking {NFA} equivalence with
  bisimulations up to congruence}}.
\newblock In: {\sl \bibinfo{booktitle}{Proc. of POPL '13}},
  \bibinfo{publisher}{ACM}, pp. \bibinfo{pages}{457--468},
  \doi{10.1145/2429069.2429124}.

\bibitemdeclare{inproceedings}{b:weighted-bisimulation}
\bibitem{b:weighted-bisimulation}
\bibinfo{author}{Michele \surnamestart Boreale\surnameend}
  (\bibinfo{year}{2009}): \emph{\bibinfo{title}{Weighted bisimulation in linear
  algebraic form}}.
\newblock In: {\sl \bibinfo{booktitle}{Proc. of CONCUR '09}},
  \bibinfo{publisher}{Springer}, pp. \bibinfo{pages}{163--177},
  \doi{10.1007/978-3-642-04081-8\_12}.
\newblock \bibinfo{note}{{LNCS} 5710}.

\bibitemdeclare{article}{Classen:2013:FTS}
\bibitem{Classen:2013:FTS}
\bibinfo{author}{Andreas. \surnamestart Classen\surnameend},
  \bibinfo{author}{Maxime \surnamestart Cordy\surnameend},
  \bibinfo{author}{Pierre-Yves \surnamestart Schobbens\surnameend},
  \bibinfo{author}{Patrick \surnamestart Heymans\surnameend},
  \bibinfo{author}{Axel \surnamestart Legay\surnameend} \&
  \bibinfo{author}{Jean-Fran{\c{c}}ois \surnamestart Raskin\surnameend}
  (\bibinfo{year}{2013}): \emph{\bibinfo{title}{Featured Transition Systems:
  Foundations for Verifying Variability-Intensive Systems and Their Application
  to {LTL} Model Checking}}.
\newblock {\sl \bibinfo{journal}{IEEE Trans. Softw. Eng.}}
  \bibinfo{volume}{39}(\bibinfo{number}{8}), pp. \bibinfo{pages}{1069--1089},
  \doi{10.1109/TSE.2012.86}.

\bibitemdeclare{inproceedings}{DBLP:conf/icse/CordyCPSHL12}
\bibitem{DBLP:conf/icse/CordyCPSHL12}
\bibinfo{author}{Maxime \surnamestart Cordy\surnameend},
  \bibinfo{author}{Andreas \surnamestart Classen\surnameend},
  \bibinfo{author}{Gilles \surnamestart Perrouin\surnameend},
  \bibinfo{author}{Pierre-Yves \surnamestart Schobbens\surnameend},
  \bibinfo{author}{Patrick \surnamestart Heymans\surnameend} \&
  \bibinfo{author}{Axel \surnamestart Legay\surnameend} (\bibinfo{year}{2012}):
  \emph{\bibinfo{title}{Simulation-based abstractions for software product-line
  model checking}}.
\newblock In: {\sl \bibinfo{booktitle}{Proc. of ICSE '12}}, pp.
  \bibinfo{pages}{672--682}, \doi{10.1109/ICSE.2012.6227150}.

\bibitemdeclare{book}{opac-b1085541}
\bibitem{opac-b1085541}
\bibinfo{author}{Raymond~A. \surnamestart Cuninghame-Green\surnameend}
  (\bibinfo{year}{1979}): \emph{\bibinfo{title}{Minimax algebra}}.
\newblock \bibinfo{series}{Lecture Notes in Economics and Mathematical
  Systems}, \bibinfo{publisher}{Springer-Verlag},
  \doi{10.1007/978-3-642-48708-8}.

\bibitemdeclare{book}{Hensel:lifting}
\bibitem{Hensel:lifting}
\bibinfo{author}{Abhijit \surnamestart Das\surnameend} \&
  \bibinfo{author}{C.~E.~Veni \surnamestart Madhavan\surnameend}
  (\bibinfo{year}{2009}): \emph{\bibinfo{title}{{P}ublic-{K}ey {C}ryptography:
  {T}heory and {P}ractice}}.
\newblock \bibinfo{publisher}{{P}earson {E}ducation}.
\newblock \bibinfo{note}{{p}p. 295-296}.

\bibitemdeclare{book}{dp:lattices-order}
\bibitem{dp:lattices-order}
\bibinfo{author}{Brian~A. \surnamestart Davey\surnameend} \&
  \bibinfo{author}{Hilary~A. \surnamestart Priestley\surnameend}
  (\bibinfo{year}{2002}): \emph{\bibinfo{title}{Introduction to lattices and
  order}}.
\newblock \bibinfo{publisher}{Cambridge University Press},
  \doi{10.1017/CBO9780511809088}.

\bibitemdeclare{book}{hwaDKV}
\bibitem{hwaDKV}
\bibinfo{editor}{Manfred \surnamestart Droste\surnameend},
  \bibinfo{editor}{Werner \surnamestart Kuich\surnameend} \&
  \bibinfo{editor}{Heiko \surnamestart Vogler\surnameend}, editors
  (\bibinfo{year}{2009}): \emph{\bibinfo{title}{Weighted Automata Algorithms}}.
\newblock \bibinfo{publisher}{Springer}, \doi{10.1007/978-3-642-01492-5\_6}.

\bibitemdeclare{inproceedings}{Esparza2014}
\bibitem{Esparza2014}
\bibinfo{author}{Javier \surnamestart Esparza\surnameend},
  \bibinfo{author}{Michael \surnamestart Luttenberger\surnameend} \&
  \bibinfo{author}{Maximilian \surnamestart Schlund\surnameend}
  (\bibinfo{year}{2014}): \emph{\bibinfo{title}{FPsolve: A Generic Solver for
  Fixpoint Equations over Semirings}}.
\newblock In: {\sl \bibinfo{booktitle}{Proc. of CIAA '14}},
  \bibinfo{publisher}{Springer}, pp. \bibinfo{pages}{1--15},
  \doi{10.1007/978-3-319-08846-4\_1}.
\newblock \bibinfo{note}{{LNCS} 8587}.

\bibitemdeclare{inproceedings}{kmoww:language-equ-prob}
\bibitem{kmoww:language-equ-prob}
\bibinfo{author}{Stefan \surnamestart Kiefer\surnameend},
  \bibinfo{author}{Andrzej~S. \surnamestart Murawski\surnameend},
  \bibinfo{author}{Joel \surnamestart Ouaknine\surnameend},
  \bibinfo{author}{Bj{\"o}rn \surnamestart Wachter\surnameend} \&
  \bibinfo{author}{James \surnamestart Worrell\surnameend}
  (\bibinfo{year}{2011}): \emph{\bibinfo{title}{Language Equivalence for
  Probabilistic Automata}}.
\newblock In: {\sl \bibinfo{booktitle}{Proc. of CAV '11}},
  \bibinfo{publisher}{Springer}, pp. \bibinfo{pages}{526--540},
  \doi{10.1007/978-3-642-22110-1\_42}.
\newblock \bibinfo{note}{{LNCS} 6806}.

\bibitemdeclare{inproceedings}{KK14}
\bibitem{KK14}
\bibinfo{author}{Barbara \surnamestart K{\"o}nig\surnameend} \&
  \bibinfo{author}{Sebastian \surnamestart K{\"u}pper\surnameend}
  (\bibinfo{year}{2014}): \emph{\bibinfo{title}{Generic Partition Refinement
  Algorithms for Coalgebras and an Instantiation to Weighted Automata}}.
\newblock In: {\sl \bibinfo{booktitle}{Proc. of TCS '14}},
  \bibinfo{series}{IFIP AICT}, \bibinfo{publisher}{Springer}, pp.
  \bibinfo{pages}{311--325}, \doi{10.1007/978-3-662-44602-7\_24}.
\newblock \bibinfo{note}{{LNCS} 8705}.

\bibitemdeclare{article}{KK16}
\bibitem{KK16}
\bibinfo{author}{Barbara \surnamestart K{\"{o}}nig\surnameend} \&
  \bibinfo{author}{Sebastian \surnamestart K{\"{u}}pper\surnameend}
  (\bibinfo{year}{2016}): \emph{\bibinfo{title}{A generalized partition
  refinement algorithm, instantiated to language equivalence checking for
  weighted automata}}.
\newblock {\sl \bibinfo{journal}{Soft Computing}}, pp. \bibinfo{pages}{1--18},
  \doi{10.1007/s00500-016-2363-z}.

\bibitemdeclare{article}{Krob94theequality}
\bibitem{Krob94theequality}
\bibinfo{author}{Daniel \surnamestart Krob\surnameend} (\bibinfo{year}{1994}):
  \emph{\bibinfo{title}{The equality problem for rational series with
  multiplicities in the tropical semiring is undecidable}}.
\newblock {\sl \bibinfo{journal}{International Journal of Algebra and
  Computation}} \bibinfo{volume}{4}(\bibinfo{number}{3}), pp.
  \bibinfo{pages}{405--425}, \doi{10.1142/S0218196794000063}.

\bibitemdeclare{article}{doi:10.1142/S0129054110007192}
\bibitem{doi:10.1142/S0129054110007192}
\bibinfo{author}{Orna \surnamestart Kupferman\surnameend} \&
  \bibinfo{author}{Yoad \surnamestart Lustig\surnameend}
  (\bibinfo{year}{2010}): \emph{\bibinfo{title}{Latticed Simulation Relations
  and Games}}.
\newblock {\sl \bibinfo{journal}{International Journal of Foundations of
  Computer Science}} \bibinfo{volume}{21}(\bibinfo{number}{02}), pp.
  \bibinfo{pages}{167--189}, \doi{10.1142/S0129054110007192}.

\bibitemdeclare{mastersthesis}{m:gen-werkzeug-sprachaequ}
\bibitem{m:gen-werkzeug-sprachaequ}
\bibinfo{author}{Christine \surnamestart Mika\surnameend}
  (\bibinfo{year}{2015}): \emph{\bibinfo{title}{{Ein generisches Werkzeug
  f{\"u}r Sprach{\"a}quivalenz bei gewichteten Automaten}}}.
\newblock Master's thesis, \bibinfo{school}{Universit\"at Duisburg-Essen}.

\bibitemdeclare{book}{p:prob-automata}
\bibitem{p:prob-automata}
\bibinfo{author}{A.~\surnamestart Paz\surnameend} (\bibinfo{year}{1971}):
  \emph{\bibinfo{title}{Introduction to Probabilistic Automata}}.
\newblock \bibinfo{publisher}{Academic Press, New York}.

\bibitemdeclare{article}{DBLP:journals/iandc/Schutzenberger61b}
\bibitem{DBLP:journals/iandc/Schutzenberger61b}
\bibinfo{author}{Marcel-Paul \surnamestart Sch{\"u}tzenberger\surnameend}
  (\bibinfo{year}{1961}): \emph{\bibinfo{title}{On the Definition of a Family
  of Automata}}.
\newblock {\sl \bibinfo{journal}{Information and Control}}
  \bibinfo{volume}{4}(\bibinfo{number}{2--3}), pp. \bibinfo{pages}{245--270},
  \doi{10.1016/s0019-9958(61)80020-x}.

\bibitemdeclare{article}{t:polynomial-equ-prob-automata}
\bibitem{t:polynomial-equ-prob-automata}
\bibinfo{author}{Wen-Guey \surnamestart Tzeng\surnameend}
  (\bibinfo{year}{1992}): \emph{\bibinfo{title}{A polynomial-time algorithm for
  the equivalence of probabilistic automata}}.
\newblock {\sl \bibinfo{journal}{SIAM Journal on Computing}}
  \bibinfo{volume}{21}(\bibinfo{number}{2}), pp. \bibinfo{pages}{216--227},
  \doi{10.1137/0221017}.

\end{thebibliography}
\end{document}